\documentclass[preprint2]{aastex}
\usepackage{times}
\shorttitle{ Non-Keplerian Dynamics }
\shortauthors{ FABRYCKY}
\usepackage{natbib}
\begin{document}
\title{\textbf{\LARGE Non-Keplerian Dynamics}}

\author{\textbf{\large Daniel C. Fabrycky\altaffilmark{1}}}
\affil{{\slshape {\bfseries Harvard-Smithsonian Center for Astrophysics} } }

\altaffiltext{1}{Michelson Fellow, daniel.fabrycky@gmail.com}

\begin{abstract}
\begin{list}{ } {\rightmargin 1in}
%{\leftmargin -1in}
\baselineskip = 11pt
%rule{4.75in}{0.5pt}
%\vskip 1pt
\parindent=1pc
{\small Exoplanets are often found with short periods or high eccentricities, and multiple-planet systems are often in resonance.  They require dynamical theories that describe more extreme motions than those of the relatively placid planetary orbits of the Solar System.  We describe the most important dynamical processes in fully-formed planetary systems and how they are modeled.  Such methods have been applied to detect the evolution of exoplanet orbits in action and to infer dramatic histories from the dynamical properties of planetary systems.
 \\~\\~\\~}%leave this in to get the correct vertical space after the abstract
 
\end{list}
\end{abstract}
% \keywords{ celestial mechanics --- planetary systems }

\section{INTRODUCTION} \label{sec:intro}

     After a planet has formed via giant impacts and has outlasted migration torques from the gaseous disk, perils still await.  Dynamical instabilities among the planets of the system, long-term orbital changes due to a companion star, and tidal interactions with the host star could all eject the planet or toss it into the host star.  None of these would threaten planets if they continued on Keplerian orbits (as described in chapter 1 of this volume), in which the planet and star are considered as point masses, orbiting each other according to Newton's approximate theory of gravity, in isolation from other bodies.  To model these more interesting interactions, we explore non-Keplerian dynamics, an extension of orbital theory to (1) the astrophysical two-body problem in more detail, including non-spherical bodies and relativity theory, and (2) true planetary systems: systems with more than just one planet and one star.  

     Although most dynamical concepts were originally designed to describe the Solar System, the discovery of exoplanets has channeled research in new directions.  Previously most analytical and even numerical techniques required certain quantities (mass ratios, eccentricities, mutual inclinations) to be small.  These approximations are appropriate for applications in the Solar System, but more general methods are needed to describe the bewildering variety of exoplanetary systems.  Here we describe these methods: both new directions of analytic theory and an introduction to numerical work.  
     
 Orbital motion departs from fixed Keplerian ellipses on a wide range of timescales, from orbital periods to stellar lifetimes.   These variations have a wide range of magnitudes, from slight deflections to complete reorientations and ejections.  Slight deflections, due to planets passing one another, may be too small to detect, apart from sensitive transit timing measurements.  Secular interactions and dissipative effects may be too slow to detect in individual systems, but they may be inferred by statistical studies of populations.   Ejections may be too rare to see in action, but they are expected from numerical modeling.  Resonant interactions, however, can produce substantially non-Keplerian motion over a timescale suitable for observation, and resonant systems can stably persist for the star's entire lifetime.  
    
The plan of this chapter is as follows.  In \S\ref{sec:eqconc}, we begin with equations of motion for planets that include tidal distortion, rotational oblateness, effects of relativity, and gravitational interactions among planets.  The framework for solving these equations is presented, including introductions to both coordinate systems and numerical algorithms.  In \S\ref{sec:phenom}, the phenomena which arise from these equations are discussed.  We emphasize phenomena which are important for known systems, as well as types of orbits that could exist but are not yet observed.  Next, in \S\ref{sec:highlights}, we turn to observational highlights which have used and continue to challenge these concepts, showing how theories and systems in nature have enjoyed a symbiotic relationship.  Finally, \S\ref{sec:future} discusses observations that would be particularly useful for constraining theories of dynamical evolution.  With this structure, some individual topics in dynamics reappear in several different sections; so we provide Table~\ref{tab:topics} as a topical guide.

\begin{deluxetable}{ll}
\tablecaption{Guide to Topics} \label{tab:topics}
\tabletypesize{\normalsize}
\tablewidth{0pt}
\tablehead{
\colhead{Topic} &
\colhead{Sections} 
}
\startdata
Single-planet orbital evolution & \\
\hspace{0.2 in}	periastron advance & \ref{sec:periad}  \\
\hspace{0.2 in}	tides   & \ref{sec:tidaldiss}  \\
\hspace{0.2 in} miscellaneous & \ref{sec:2bodymisc} \\
\hline
Few-body orbital interactions &   \\
\hspace{0.2 in}	short-term orbit fluctuations & \ref{sec:types}, \ref{sec:arch}   \\
\hspace{0.2 in}		resonances& \ref{sec:types}, \ref{sec:gj876}, \ref{sec:pulsar}, \ref{sec:ttv} \\
\hspace{0.2 in}		secular effects& \ref{sec:types}, \ref{sec:advanced}, \ref{sec:seclock}, \ref{sec:kozai}  \\
\hspace{0.2 in}		chaos& \ref{sec:chaos}  \\
\hline
Dynamical niches & \\
\hspace{0.2 in}		resonance protection& \ref{sec:types}, \ref{sec:gj876}    \\
\hspace{0.2 in}		satellites& \ref{sec:otherconfig}, \ref{sec:ttv}    \\
\hspace{0.2 in}		Trojans and horseshoes& \ref{sec:otherconfig}, \ref{sec:ttv}    \\
\hspace{0.2 in}		interlocking orbits& \ref{sec:advanced}    \\
\hspace{0.2 in}		habitable zones& \ref{sec:otherconfig}\\
\hline
Data Analysis & \\
\hspace{0.2 in}		astrometry and direct imaging& \ref{sec:nbodyfit}    \\
\hspace{0.2 in}		radial velocity& \ref{sec:nbodyfit}    \\
\hspace{0.2 in}		transits& \ref{sec:ttvcalc}, \ref{sec:ttv}, \ref{sec:arch}  \\
\hline
Individual Systems & \\
\hspace{0.2 in}		GJ 876& \ref{sec:gj876}     \\
\hspace{0.2 in}		Pulsar 1257+12& \ref{sec:pulsar}    \\
\hspace{0.2 in}		16 Cyg B & \ref{sec:otherconfig}, \ref{sec:kozai}  \\
\hspace{0.2 in}		HD 80606 & \ref{sec:advanced}, \ref{sec:kozai} \\
\hspace{0.2 in} 	resonant systems & Table~\ref{tab:res}
\enddata
\end{deluxetable}

\section{ EQUATIONS OF MOTION AND NUMERICAL METHODS} \label{sec:eqconc}

\subsection{Astrophysical Two-body Problem} \label{sec:2body}
First, let us examine the astrophysical two-body problem including non-Keplerian effects.  The masses of the star and planet are $m_\star$ and $m_p$, respectively, and the orbital elements refer to the displacement vector $\mathbf{r}$ (of magnitude $r$) of the planet relative to the star, as in Chapter 1 of this volume.  The equation of motion is:
\begin{equation}
\mathbf{\ddot{r}} = - G(m_\star+m_p) \frac{\mathbf{r}}{r^3} + \mathbf{f},  \label{eq:eom}
\end{equation}
where $f$ is a force other than that of mutual gravity of point masses, so $\mathbf{f}=0$ yields Keplerian motion (Chapter 1, equation 4).

\subsubsection{Relativistic effects} \label{sec:greffects}

The lowest-order post-Newtonian effects may be implemented using the force:
\begin{eqnarray}
\mathbf{f}_{{\rm GR}} &=& -\frac{G(m_\star + m_p)}{r^2 c^2}\times \Big( - 2(2-\eta) \dot{r}\dot{\mathbf{r}} \nonumber \\ 
	& & + \Big{[} (1+3\eta)\dot{\mathbf{r}} \cdot \dot{\mathbf{r}} - \frac{3}{2} \eta \dot{r}^2  \\ \nonumber
	& & - 2 (2 + \eta)\frac{G(m_\star + m_p)}{r} \Big] \hat{\mathbf{r}} \Big) \label{eq:kidder}
\end{eqnarray}
(where $\eta = m_\star m_p / (m_\star + m_p)^2$; \citealt{1995K,2002ML}) for $\mathbf{f}$ in equation~(\ref{eq:eom}).  We will examine one dynamical consequence of this force, apsidal motion, in \S\ref{sec:periad}.  Alternatively, a potential that mimics lowest-order relativistic effects, which is especially suitable for analysis within a Hamiltonian framework, was given by \cite[eq.~31]{1992ST}.

\subsubsection{Effects of non-spherical bodies}
\label{sec:tidaleff}

The orbital parameters change when we consider the bodies not as point masses, but as physical objects capable of distortion and internal energy dissipation.  Tidal effects on the orbit become more and more pronounced as two gravitating bodies of finite extent get closer to one another.  Tides have apparently caused many exoplanet orbits to become circular within about $0.1$~AU of their main-sequence stars.  

The tidal force of each body distorts the potential energy surfaces of its companion.  For stars and gaseous planets in hydrostatic equilibrium, the surfaces of constant density will settle to these new equipotential surfaces, and the distortion of the body itself modifies them further, until a self-consistent solution is obtained.  For solid planets, the rigidity of the material also comes into play.  These properties determine the planet's tidal deformability and are summarized by a Love number $k_L$ \citep{1911L}, which is the amplitude ratio of the quadrupolar potential due to the deformed body to the tidal potential imposed on the body, evaluated at the surface of the body; a table of typical values of $k_L$ may be found in \cite{2004ML}.  Below we shall also denote the Love number of the star as $k_{L,\star}$, which is \emph{twice} the apsidal motion constant, the conventional parameter in the literature regarding eclipsing binaries \citep{1939S}.

First, the star raises a tidal bulge of the planet with a size $\propto r^{-3}$.  This bulge creates its own external field which falls off like $r^{-3}$.  Acting back on the star, these radial scalings combine to augment the radial gravitational force (per unit mass) with a term having steep radial dependence: 
\begin{equation}
\mathbf{f}_{\rm T} = - 3 k_L \frac{G m_\star^2}{m_p} \frac{R_p^5}{r^7} \mathbf{\hat{r}} . \label{eq:planettide}
\end{equation}

Second, the rotation of the star causes it to become oblate.  Its degree of oblateness depends on the square of the stellar angular rotation rate $\Omega_\star$ divided by its break-up angular rate, $\sqrt{G m_\star / R_\star^3}$.   The quadrupolar potential outside of the star scales as $r^{-3}$.  Thus the figure of the rotating star induces an extra force
\begin{equation}
\mathbf{f}_{\rm R} = - \onehalf k_{L,\star} \Omega_\star^2 \frac{R_\star^5}{r^4} \mathbf{\hat{r}}. \label{eq:stellaroblate}
\end{equation}

The forces of the foregoing two subsections may be included in numerical integrations to simulate more realistic orbital behavior than Keplerian ellipses.  They may also be used in analytic calculations to determine orbit changes on a longer timescale, results of which we quote in \S\ref{sec:twobodeff}.

\subsection{ N-body equations and coordinates}  \label{sec:nbodyeqn}

When a star hosts N ($>1$) planets, gravitational interactions among the planets can affect their orbits in complex ways.  Neglecting the effects discussed in \S\ref{sec:2body}, the equation of motion of planet planet $i$ (of mass $m_i$) is:
\begin{equation}
\mathbf{\ddot{r}}_i = - G(m_0+m_i) \frac{\mathbf{r}_i}{r_i^3} + G \sum_{j=1;j\neq i}^{N} m_j \Big( \frac{\mathbf{r}_j - \mathbf{r}_i}{|\mathbf{r}_j - \mathbf{r}_i|^3} - \frac{\mathbf{r}_j}{ r_j^3} \Big) ,  \label{eq:eomnpl}
\end{equation}
where each of the coordinates $\mathbf{r}_i$ is referred to the central star, of mass $m_0 \equiv m_\star$.  The interaction terms in the sum over each of the other planets are the \emph{direct} gravitational force (first term) and the \emph{indirect} effective force due to the bodies causing the star, and thus the reference frame, to accelerate (second term).  In numerical work, these $N$ second-order differential equations are most often transformed into a system of $2N$ first-order differential equations in the quantities $\mathbf{r}_i$ and $\mathbf{\dot{r}}_i$.  In response to the motions of the planets, the star's position and velocity with respect to the barycenter of the system (the center of mass) are:
\begin{eqnarray}
\mathbf{R}_0 = - \Big( \sum_{i=1}^N m_i \mathbf{r}_i \Big) / \Big(\sum_{i=0}^N m_i \Big) \label{eq:astropos} \\
\mathbf{\dot{R}}_0 = - \Big( \sum_{i=1}^N m_i \mathbf{\dot{r}}_i \Big) / \Big(\sum_{i=0}^N m_i \Big), \label{eq:rv}
\end{eqnarray}
which are useful for self-consistent fits of data (see \S\ref{sec:nbodyfit}).

The coordinates of equation~(\ref{eq:eomnpl}) are called astrocentric, because they refer each planet to the (moving) position of the star.  There are several other possible coordinate systems---barycentric, Jacobian, or Poincar{\'e} coordinates---which are defined as follows.  Barycentric coordinates refer all positions and velocities to the center of mass, and the only force is Newton's attractive force between bodies (no indirect term).  Jacobian coordinates are hierarchical, in which the positions and velocities of each of the planets is referred to the center of mass of all interior bodies.  Poincar{\'e} coordinates [\citealt{1995LR}; also called democratic heliocentric coordinates \citep{1998D} or canonical heliocentric \citep[pg. 435]{2001SW}], on the other hand, take the positions of the planets to be astrocentric, but take the velocities of the planets to be barycentric.  

The various coordinate systems have different strengths and weaknesses, in both numerical integration and in analytical studies, as follows.  

Astrocentric coordinates have a few drawbacks.  In numerical integrations, all planets must be integrated with timesteps smaller than the innermost planet's period, because of the high-frequency forcing it introduces.  Also, for theoretical studies in a Hamiltonian framework, the astrocentric coordinates are rather cumbersome \citep{2007Beau}.  

Although barycentric coordinates are very simple, they do not take advantage of the fact that the center of mass is invariant.  Therefore the position of the star must either be numerically integrated or updated at each timestep according to equation~(\ref{eq:astropos}).  A reduction by those 3 degrees of freedom is achieved using Jacobian or Poincar{\'e} coordinates.  

\cite{2003LP} have advocated using Jacobian coordinates, instead of astrocentric coordinates, when turning the solutions of multiple Keplerian radial velocity fits into a self-consistent N-body realization (\S\ref{sec:nbodyfit}) .  Particularly for systems that are hierarchical, these are the coordinates in which the planets perturb each other minimally on orbital timescales, so the independent-Keplerian model is satisfied best.    Once an integration is set up in Jacobian coordinates, one might want to do the integration in astrocentric coordinates, so that, e.g., transit times of the outer planet can be easily calculated (\S\ref{sec:ttvcalc}).  The hierarchical structure of Jacobian coordinates is not always physically well-motivated, e.g., for systems with Trojan orbits.  Most authors take the relevant mass binding the $j^{th}$ planet of mass $m_j$ to the mass interior to it $m_{\rm interior,j}$ to be $m_j+m_{\rm interior,j}$ (in generalization of the mass term eq.~\ref{eq:eom}), but in symplectic integrations (described in \S\ref{sec:numtech}) it can be more efficient to use $\tilde{M}_j \equiv m_\star m_{\rm interior,j}/m_{\rm interior,j-1}$ \citep{1991HW,1999MD}.  

Poincar{\'e} coordinates have positions and velocities with different origins, so they are not conceptually based on elliptical motion.  However, analytic studies can be cleaner, as the resulting Hamiltonian has rather clear symmetry.  

For each coordinate system, each planet has ``osculating'' orbital elements which correspond to the orbit the planet would have if all the interaction forces are ignored.  For instance, orbital elements based on the astrocentric coordinates have a simple physical interpretation: if $(N-1)$ planets of the system suddenly disappeared, a single planet would be left orbiting the star with those orbital elements.  Osculating elements have a draw-back, which is that the physical, observed properties, which include the continual perturbations of the other bodies, do not correspond to the osculating value.  For instance, the orbital period observed with transits or radial velocities is not the osculating period in an astrocentric coordinate system (for a conversion, see \citealt{2005F}). 

For explicit forms of these coordinate systems, and a fuller discussion, see \cite{2007Beau} and \cite{2002Morbi}.  

    \subsection{ Numerical integration techniques } \label{sec:numtech}
     
     Here we summarize some of the numerical techniques and codes that are used in exoplanet dynamics.
     
     Currently, the fastest algorithms that are reliable for long-term studies are symplectic integrators, in which separate integrations delimiting a certain volume of phase space continue to delimit the same volume as the integration proceeds.  This property automatically respects the Hamiltonian nature of the gravitational problem; e.g., energy is not secularly lost or gained.  The \cite{1991HW} integrator is the most well-known of the symplectic integrators.  It has been supplemented with algorithms that can handle close approaches and collisions, which forms the backbone of the publicly-available codes {\slshape Mercury}\footnote{\url{http://www.arm.ac.uk/\~jec/home.html}} \citep{1999C}, {\slshape Swift}\footnote{\url{http://www.boulder.swri.edu/\~hal/swift.html}} \citep{1994LD}, and {\slshape SyMBA}\footnote{Available upon request from Hal Levison.} \citep{1998D}.  A symplectic algorithm was also built specifically to handle stellar mass companions \citep{2002C}, either with planets orbiting one star of the pair (satellite-type) or both stars (planetary-type).  The speed of symplectic integrators allowed some of the first long-term integrations of all the planets of the solar system, showing their orbits to be chaotic (e.g., \citealt{1992SW}).  High-order symplectic schemes have been used to follow the whole Solar System for its lifetime \citep{2009LG}.
     
     There are several frequently-used, but non-symplectic, algorithms.  RADAU (15) by \cite{1985E} is a very high-order method of integration by Radau quadrature --- a prescription for times to evaluate forces and weights to apply to them --- and it is included in {\slshape Mercury}.  General-purpose integrators from {\slshape Numerical Recipes} (\citealt{1992P}, e.g., the popular Burlisch-Stoer) and the GNU Scientific Library [e.g., Embedded Runge-Kutta Prince-Dormand (8/9 order), as advocated by \citealt{2004F}] have also been set to work on problems in exoplanet dynamics.  These integrators are particularly useful for short-term, high precision work.
    
     If mean motion resonances are unimportant, one can derive approximate secular equations of motion by analytically averaging each planets' gravitational effect over its orbit.  This procedure results in differential equations for the orbital elements, which can be numerically integrated much faster than the positions themselves.  \cite{2002ML} have presented equations to evolve three-planet systems, as well as separate equations for the system once the inner planet's orbit is averaged.  \cite{2001EKE} have presented equations for the secular problem for three-body systems where the semi-major axis ratio is large, and the outermost orbit dominates the angular momentum budget; this approach is applicable to planets in binaries.  Higher order expansions in the ratio of semi-major axes, with an emphasis on planetary systems, have also been derived \citep{2000F, 2003LP, 2008MG}.  Taking no restrictions on relative size of orbits or orbital elements, one can compute by brute force the time-averaged torque acting among a set of nearly-Keplerian orbits, which is called Gauss's method (e.g., \citealt{2009T}).

    Finally, a general consideration, which is surprising at first, is that codes are usually much slower at printing data to files than computing data.  For instance, if the position of the planets is printed once per orbit, the print-outs will generally completely dominate the runtime.  It is generally true that performance can be limited by the weakest link (the slowest part of the algorithm), so one must think about the program as a whole, including the reads and writes, when trying to speed up an algorithm.

\subsection{ N-body fits to data } \label{sec:nbodyfit}

	To directly detect non-Keplerian motion, a framework must be constructed for comparing the outputs of numerical algorithms to data.  We describe the conventional approach here.

         The astrocentric coordinates for planets are $\mathbf{r}_i=(x_i, y_i, z_i)$; the barycentric coordinates of the planets are $\mathbf{R}_i=(X_i, Y_i, Z_i)$, and the barycentric coordinates of the host star are $\mathbf{R}_0=(X_0, Y_0, Z_0)$, which obey equation~(\ref{eq:rv}).  The standard coordinate system (e.g., Chapter 1) has the $Z$-axis pointed away from the observer.  Let $X-Y$ plane be the plane of the sky, which passes through the barycenter.  Then aligning the $X$-axis with the North direction and the $Y$-axis with the East direction forms a right-handed coordinate system with $Z$.  The ascending node $\Omega$ is measured East from North, and the argument of periastron $\omega$ is measured from the sky plane.  These coordinates are useful for fitting data from essentially all techniques (neglecting the rectilinear motion of the system as a whole): 
          \begin{enumerate}
          \item{ radial velocity of the host~$= -\dot{Z}_0$,}
          \item{ astrometric position of the host~$= (X_0, Y_0)$,  }
          \item{ transits and secondary eclipses may be modeled using the light curve equations of \cite{2002MA}, setting the projected distance between the centers, called $d$, equal to $\sqrt{x_i^2+y_i^2}$ (see also \S\ref{sec:ttvcalc} below),}
          \item{ direct images of planets have a sky-offset from their host of $(x_i, y_i)$,}
          \item{ the times of arrival (TOA) of pulses of pulsars, or phases of pulsating stars, or eclipses of close binaries, that are orbited by planets are delayed by $\delta t = -Z_0/c$, etc. }
          \end{enumerate}
          
	Therefore a common coordinate system can be used to fit different types of datasets simultaneously, as has already been done for radial velocity plus astrometry \citep{2009BS,2009WH} and for radial velocity plus transits (e.g., \citealt{2005WinnSO}).

	The values of $G$ and $m_\odot$ are not known to nearly the precision of their combination.  A result is that dynamical fits reported in the literature are typically not reproducible if the masses are reported in physical units (e.g., grams).  One solution is to report all masses in units of solar masses.  Alternatively, one could abide by the International Astronomical Union's conventional value of $k \equiv \sqrt{G m_\odot}=0.01720209895$~AU$^{3/2}$~day$^{-1}$.
	
	There are numerous methods for converging on the solution and evaluating uncertainties.  Markov Chain Monte Carlo \citep{2005Funcertain,2006F}, Levenberg-Marquardt \citep{2001LC,2001RL}, genetic algorithm \citep{2005G}, and multiplexed simulated annealing (also known as parallel tempering; \citealt{2007Gregory}) have all been used extensively, and these references describe the implementations and advantages of each method.

	The magnitude and/or timescale of perturbations scale as powers of the mass ratios, so the true planetary masses in systems with measurable non-Keplerian motion are accessible (\S\ref{sec:gj876},\ref{sec:pulsar}).  Moreover, often only a subset of orbital parameters which fit the observations actually result in long-term stability, so additional information about the system is accessible by requiring long-term stability.  This idea was pursued for the first multiplanet system by \cite{2000RL,2000SMB}, and it has been particularly useful for resonant systems; see \cite{2008Goz} for an overview.
	
\subsection{ A focus on transit variations } \label{sec:ttvcalc}
	
	Now let us focus on dynamical fits to transit data, which have the potential to discover small planets and to reveal the detailed dynamical properties of multiplanet systems (see \S\ref{sec:future}).  The data begin as flux measurements of a star as a function of time; the star is dimmed as the transiting planet covers some of the stellar surface (see chapter 4).  Encoded in this time series is information about the position of the planet relative to the star, so integrators based on astrocentric coordinates are the most natural for this application.  The data allow for precise positions and times to be observed, so high precision in coordinates and arbitrary timesteps is required.  Typically, the flux measurements are used to derive  transit parameters --- mid-transit times and durations (or impact parameters or inclinations) --- as a function of transit number, which may be compared directly to a numerical integration.
	
\begin{figure}[t]
\includegraphics[scale=0.8]{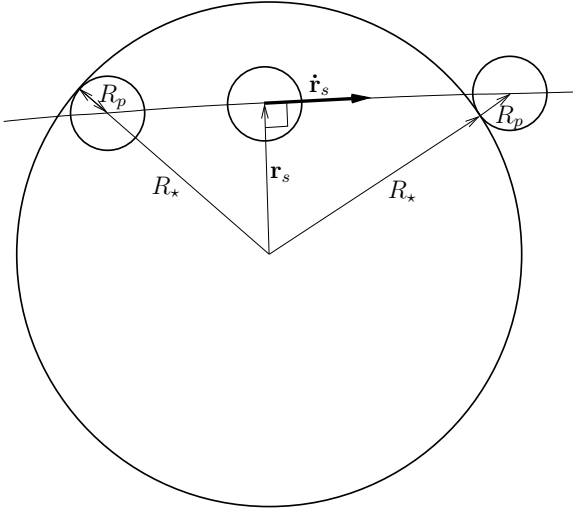}
\caption{  Diagram of a transiting planet at the contact and mid-transit points of its orbit. }
\label{fig:transit}
\end{figure}

	Theoretical mid-transit and contact times for the $i^{th}$ planet may be computed by integrating forward and backward in time by the Newton-Raphson method, seeking roots of functions of $\mathbf{r}_{s,i}\equiv(x_i,y_i)$, the relative separation vector of the planet and star on the plane of the sky; see Figure~\ref{fig:transit}.  This method is described next.  The only requirement for the following algorithm is that the sky-projected trajectory has a radius of curvature larger than $R_\star + R_p$, which is easily fulfilled in practice.  
	
	For each transit the mid-time is found first, as follows.  The first step is to advance the integrator in time to the vicinity of an observed transit, for instance, by stopping an integration once an observed transit time has been passed, or advancing the appropriate number of nominal orbital periods from a previous transit.  Mid-transit times may be found by minimizing $|\mathbf{r}_{s,i}|$.  Taking the derivative of this quantity (expressed in $x_i$ and $y_i$) and setting it to zero, we find that minimizing $|\mathbf{r}_{s,i}|$ amounts to solving
\begin{equation}
g(x_i, \dot{x}_i, y_i, \dot{y}_i) \equiv \mathbf{r}_{s,i} \cdot \mathbf{\dot{r}}_{s,i} = x_i\dot{x}_i + y_i\dot{y}_i = 0. \label{eq:midtime}
\end{equation}
The first guess at the time of transit must be within about $1/8$ of a period, or the algorithm may converge to an adjacent local maximum of $|\mathbf{r}_{s,i}|$.  The solution of equation~\ref{eq:midtime} is found by $5$-$10$ iterations of moving the integrator by
\begin{equation}
\delta t = - g \Big(\frac{\partial g}{\partial t}\Big)^{-1},
\end{equation}
where 
\begin{equation}
\frac{\partial g}{\partial t} = \dot{x}_i^2 + x_i \ddot{x}_i + \dot{y}_i^2 + y_i \ddot{y}_i
\end{equation}
according to equation~(\ref{eq:midtime}).  Once $\delta t$ is below the required accuracy, the mid-time $t_{\rm mid}$ and position $\mathbf{r}_{s,i}^{\rm mid}$ of that transit have been found.  This method is computationally considerably faster than searching for a minimum of $|\mathbf{r}_{s,i}|$ directly, as root-finding is a simpler operation than minimum-finding. 

Next, before solving for times of contact, we can determine if they exist for this particular transit.  (See chapter 4 for the definitions of points of contact and for the numbering scheme.)  If $|\mathbf{r}_{s,i}^{\rm mid}| < R_\star - R_p$, then all four times of contact exist; the trajectory is not grazing.  If $R_\star - R_p< |\mathbf{r}_{s,i}^{\rm mid}| < R_\star + R_p$, then the trajectory is grazing and only first and fourth contact exist.  If $|\mathbf{r}_{s,i}^{\rm mid}| > R_\star + R_p$, the planet is not transiting at all and searching for times of contact is a waste.  To search for times of contact, first advance the integrator to $t_{\rm mid} \mp R_\star / |\mathbf{\dot{r}}_{s,i}^{\rm mid}|$, where $-$ is taken for first and second contact, and $+$ is taken for third and fourth contact.  Next, solve the equation 
\begin{equation}
h(x_i, y_i) = |\mathbf{r}_{s,i}|^2 - (R_\star \pm R_p)^2 = 0, \label{eq:contact}
\end{equation}
where $+$ is taken for first and fourth contact, and $-$ is taken for second and third contact.  As before, this is done by iteratively driving the integrator by 
\begin{equation}
\delta t = - h \Big(\frac{\partial h}{\partial t}\Big)^{-1}, \label{eq:dth}
\end{equation}
where now
\begin{equation}
\frac{\partial h}{\partial t} = 2 x_i \dot{x}_i + 2 y_i \dot{y}_i.
\end{equation}
It is important not to start the search near mid-transit, otherwise $\frac{\partial h}{\partial t} = 2 g \approx0$ and equation~(\ref{eq:dth}) will ask for an enormous jump: preemptively advancing to near or beyond the times of contact (as above) avoids that fate.  As usual with Newton-Raphson's method, sensible bounds should be placed on the search for the root.

	After times of contact and mid-transit are computed, they need to be corrected for the finite travel time of light.  Thus the apparent time of an event depends on the distance between the bodies and the center of mass of the system, along the line of sight (see, e.g., \citealt{2007K,2009B}).  Suppose, for instance, that the ephemeris of mid-transit times for a transiting planet is well-established, and the values of eccentricity $e$ and longitude of periastron $\omega$ are known.  Then the time of secondary eclipse could be predicted, but it will actually be observed a time $\Delta t$ later due to the finite speed of light, where:
\begin{equation}
\Delta t = \frac{2 a}{c} \frac{m_\star^2 - m_p^2}{(m_\star + m_p)^2}  \frac{1-e^2}{1- e^2 \sin^2 \omega}.
\end{equation}
This equation is derived by delaying the image of each body according to its line-of-sight distance, then solving for the time at which these images cross, i.e., the mid-time of secondary eclipse as observed.  This delay can be large for transiting planets, e.g. $\Delta t \approx 160$~s for HD80606b, a systematic effect that is comparable in magnitude to the measurement error of $\sim260$~s for an individual secondary eclipse \citep{2009Laughlin}.  The reason this effect has an observable magnitude is because it originates within special relativity, and so it scales as $1/c$.  Effects of general relativity, such as those embodied in equation~(\ref{eq:kidder}) or the effect of curvature on light propagation \citep{1964S}, scale as $1/c^2$, and are thus harder to detect.

	A final step is needed to compare these theoretical event times to the observed data.  The most important effect of a second body on transit times is to change the true period (as mentioned in \S\ref{sec:nbodyeqn}), so in practice the times are scaled to match the observed period and shifted to match the observed epoch of transit.  Then variations on top of that simple linear ephemeris can be used, e.g., to detect or set limits on the presence of perturbing bodies (\S\ref{sec:ttv}).  Care needs to be taken if radial velocities are being simultaneously fit, because such data specify a particular \emph{velocity} of the star corresponding to a particular \emph{time}, whereas transit data specify a particular \emph{time} corresponding to a particular relative \emph{position} of the planet to the star.

\section{ DYNAMICAL PHENOMENA } 
\label{sec:phenom}

\subsection{ Astrophysical Two-Body Problem} \label{sec:twobodeff}

\subsubsection{ Periastron advance } \label{sec:periad}
In \S\ref{sec:2body} we encountered several extra forces which modify realistic two-body motion from Keplerian ellipses.  Let us take a perturbation theory approach, in which the force is calculated over a Keplerian orbit, to find how that orbit itself evolves \citep{1976B}.   Now that the effective force of gravity is no longer of the form $1/r^2$, the ellipse does not close, and the periastron advances, which amounts to a reorientation of the orbit within its own plane.  We shall calculate this precession rate due to the perturbing forces of relativity, tides, and rotational oblateness.  

Periastron advance is the relativistic effect which changes orbits on the shortest timescale, and averaging equation~(\ref{eq:kidder}) over an orbit, we determine its angular rate to be:
\begin{equation}
\dot{\omega}_{\rm GR} = \frac{3 G^{3/2} (m_\star+m_p)^{3/2}}{a^{5/2} c^2 (1-e^2)}. \label{eq:grprec}
\end{equation}
This effect causes an additional 43 seconds of arc per century of precession for Mercury (in addition to the precession caused by the other planets), which was the famous first hint that nature obeyed Einstein's equations.  Higher order corrections to this precession rate, precession due to the star's spin (the Lenz-Thirring effect), orbital decay due to gravitational wave emission, and other relativistic effects are all negligible for exoplanets.

The force due to tidal distortion of the planet (eq.~[\ref{eq:planettide}]) causes apsidal motion at the rate:
\begin{equation}
\dot{\omega}_{\rm T} = \frac{15}{2} n k_L \frac{m_\star}{m_p} \frac{ 1 + (3/2)e^2 + (1/8)e^4 }{(1-e^2)^5} \Big(\frac{R_p}{a} \Big)^5,
\end{equation}
where $n \equiv 2\pi/P$ is the mean motion \citep{1939S}.  This effect is generally much bigger than that of the tide raised on the star by the planet, and for hot Jupiters with periods less than three days, it can dominate all other precessional effects \citep{2009RW}.  For physically smaller planets, and for Jupiter-type planets with periods $\gtrsim 3$~days, relativistic precession (eq.~[\ref{eq:grprec}]) typically dominates.  

Rotational distortion gives rise to a force (eq.~[\ref{eq:stellaroblate}]) that causes a periastron advance rate:
\begin{equation}
\dot{\omega}_{\rm R} = \frac{n k_{L,\star}}{2} \frac{1+m_p/m_\star}{(1-e^2)^2} \Big(\frac{\Omega_\star}{n} \Big)^2 \Big(\frac{R_\star}{a} \Big)^5. \label{eq:odspin}
\end{equation}  
Around fast-rotating and large stars (i.e., young or early-type), this effect can dominate the others.  If the stellar spin is misaligned with the orbit by an angle $\psi$, equation~(\ref{eq:odspin}) requires an extra term $(5 \cos^2 \psi - 1)/4$; for spin-orbit angles satisfying $63.4^\circ<\psi< 116.6^\circ$, the apsidal motion is retrograde.  With spin-orbit misalignment, the nodal angle also precesses; equations for the coupled spin and orbital motion are given by \cite{2001EKE}.  

Of course, the star has a tidal bulge and the planet has a rotational bulge as well, but these never contribute substantially to the total precession.

\subsubsection{Tidal dissipation} \label{sec:tidaldiss}

Tidal energy is converted to heat when a tidal bulge rotates through a body or varies in amplitude, due to the material's resistance to shearing motion.  First, the dissipative torque changes the rotation of the planet to a rate at which the time average of that torque vanishes.  At this spin rate the time average of the shear, and the energy dissipation rate, is minimized.  In a fixed, circular orbit, the spin angular velocity equals the orbital angular velocity and the obliquity is zero, so in the frame corotating with the perturber, the tide is no longer time variable, stopping energy loss.  In an eccentric orbit, the spin will either settle at a pseudo-synchronous state \citep{1965PG,1981Hut, 2007Levrard}, or be trapped in a spin-orbit resonance (of which Mercury is the prototype); the latter is only possible if the body has a permanent quadrupole moment due to its rigidity, and is therefore not expected for gas giants.  For rocky planets with dynamically-important atmospheres (of which Venus is the prototype), the picture can be qualitatively different, including up to four stable rotation states \citep{2008C}.

On a longer timescale, the eccentricity damps.  The correlation between eccentricity and orbital distance (or period) is the main constraint on tidal theory for exoplanets (see chapter 9).  This damping can in principle be due to either dissipation in the planet or the star.  If dissipation in the star is important, eventually the planetary orbit will decay into the star (e.g., \citealt{1996R, 2009J, 2009BO}).  If dissipation in the planet is important, then it may have ingested more tidal energy than its own binding energy.  In that case, gas giants could inflate or even disrupt \citep{2003GLB}, and such heating on terrestrial planets would have significant geophysical consequences \citep{2008W}.  

The physical causes of tidal damping for giant planets are still poorly known; as yet no first-principle theory is efficient enough to damp the eccentricity of hot Jupiters, or to generate the inferred histories of satellite systems around the four Solar System giants.  \cite{2000Lin} and \cite{2004OL} discuss these matters and review the literature.  In the absence of such a theory, a phenomenological approach has gained currency \citep{1966GS}.  A fraction $1/Q$ of the tidal energy is dissipated per tidal forcing cycle (or per orbit, depending on the author).  This allows differential equations for tidal damping to be derived, in which damping times scale with $Q$ \citep{2002ML,2008Mats}.  Empirical constraints on $Q$ for close-in gas giants have been worked out \citep{2003Wu,2008Jack, 2008Mats}.  

\subsubsection{Miscellaneous orbital evolution}
\label{sec:2bodymisc}

There are numerous other effects that can modify a planet's orbit about its star.  Here we simply list some of these effects, referring the reader to work that describes them in detail.

Close in to the star, the planet may be tidally stripped of mass.  As the mass leaves the planet, it applies a torque on its orbit.  The reaction of the orbit has been calculated for circular orbits as the planet finishes migration due to a gas disk \citep{1998T}, for moderate eccentricities as the planet tidally circularizes and perhaps inflates \citep{2003GLB}, and for eccentricities near 1 when the planet is shot near the star by either a dynamical instability or a chance flyby \citep{2005FRW}.  

Once close to the star, the planet's atmosphere absorbs and reradiates photons in preferential directions, which can lead to at most a $5\%$ change in semi-major axis --- enough to influence resonant configurations with more distant planets \citep{2008F}.  

A planet or planets may  scatter and eject a sea of small bodies (planetesimals) after the main formation phase, which leads to planetary migration.  This effect was first worked out for the giant planets of the Solar System \citep{1984FI, 1993M, 1995M}, and has since been applied to exoplanets \citep{1998M,2007M,2008T}.  For more on migration, particularly in a gas disk, see chapter 14.

Finally, far from the host star, passing stars may perturb planetary orbits (see, e.g., \citealt{2009S}).

\subsection{ Short-period, secular, and resonant interactions }  \label{sec:types}

\begin{figure}
\includegraphics[scale=0.8]{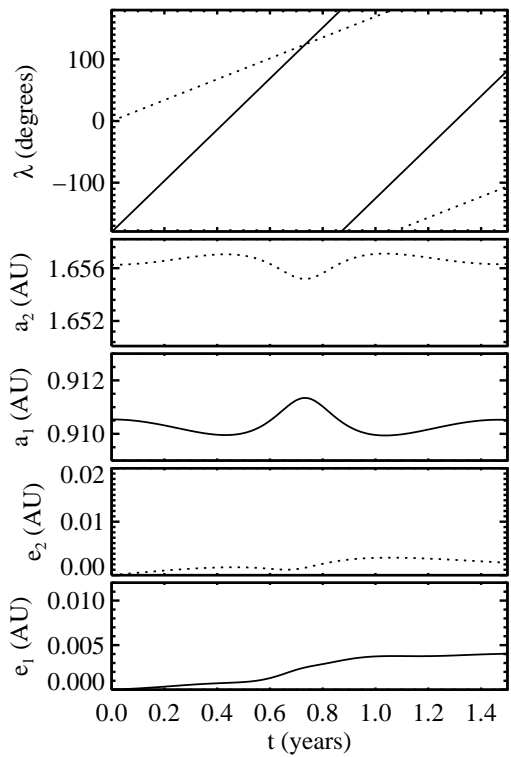}
\caption{  Orbital element changes on the timescale of conjunctions, in a hypothetical system.  The stellar mass is $m_\odot$, both planets have mass $10^{-3}m_\odot$, their orbits are coplanar and initially circular, and they start on opposite sides of the star.  Here and elsewhere, planets are numbered by increasing semi-major axis.  \emph{top panel}: the mean longitudes of each planet (the two lines cross at conjunction), \emph{second and third panels}: the planets' semi-major axes, which vary symmetrically about the conjunction, \emph{fourth and fifth panels}: the planetary eccentricites, which receive a small kick.  }
\label{fig:conjunction}
\end{figure}

The interaction terms in the equations of motion~(eq.~[\ref{eq:eomnpl}]) lead to all the interesting behavior of N-planet systems, and here we show heuristically how short-period, secular, and resonant behaviors arise.  (For a traditional expansion in terms of orbital elements, using the so-called disturbing function, see \citealt{1999MD}.)  Figure~\ref{fig:conjunction} shows how the Jacobian orbital elements of two planets, initially on circular orbits, evolve as the planets move from opposition (being on opposite sides of the star), through conjunction (lined up with the star on the same side), then back to opposition.  As the planets approach each other, the mutual gravity moves them onto slightly different orbits.  At conjunction, the inner planet has been torqued forward, to a more distant, slower orbit; the outer planet has been torqued backward, to a closer-in, faster orbit.  As the planets move through and recede from conjunction, these orbit changes are mostly reversed.  

Now let us introduce moderate eccentricity to the orbits, and follow the system for many conjunctions.  Due to the eccentricity, the paths of the planets at various conjunctions are either converging or diverging, and the changes in orbital elements on either side of conjunction do not cancel as completely.  At a single conjunction, this causes the orbits to transfer energy (the semi-major axes change) and angular momentum (the eccentricities and orbit orientations change).  After multiple conjunctions, the behavior of the system depends on whether the periods are near a ratio of small integers.

\begin{figure}
\includegraphics[scale=0.8]{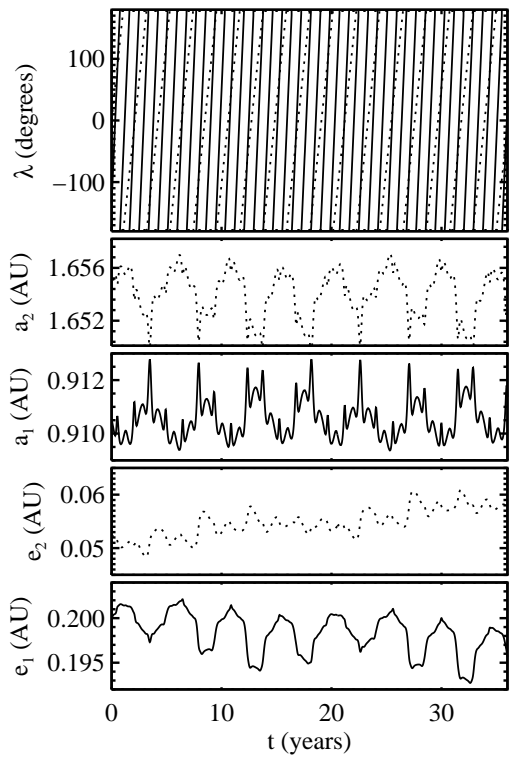}
\caption{  Orbital element changes after many conjunctions.  The hypothetical system is the same as in Figure~\ref{fig:conjunction}, except both planets start with eccentricity, with the inner planet's argument of pericenter $45^\circ$ ahead of the outer planet's.  The vertical axis on each panel has the same scaling as the corresponding panel in Figure~\ref{fig:conjunction}, emphasizing that the semi-major axes experience no net drift, but the eccentricities do. }
\label{fig:sec}
\end{figure}

First, consider Figure~\ref{fig:sec}, which shows a hypothetical system with a period ratio far from any ratio of small integers.  Because of this property, the conjunctions sample all parts of both orbits rather equally.  The semi-major axes exhibit no long-term changes, which means the energy of each orbit is conserved.  However, the angular momentum of each orbit is exchanged on long timescales, resulting in eccentricity variations.  One way of seeing why this happens is by considering the time average of a planet over its orbit, so that its gravitational effect is that of an elliptical wire weighted inversely to the Keplerian velocity at each position.  Each of the planets respond to the other planets as if they were such rings.  Because the potential from such a ring is not time dependent, it produces a conservative force, and no energy can be exchanged: semi-major axes and periods may not change.  However, the lopsided rings do torque each other, and this corresponds to angular momentum (and thus eccentricity and inclination) changes.  For instance, two eccentric, coplanar planets will undergo periodic oscillations in eccentricity which are $180^\circ$ out of phase from each other.  For more on this topic, called called secular evolution, see \S\ref{sec:advanced} and \S\ref{sec:seclock}.

\begin{figure}
\includegraphics[scale=0.8]{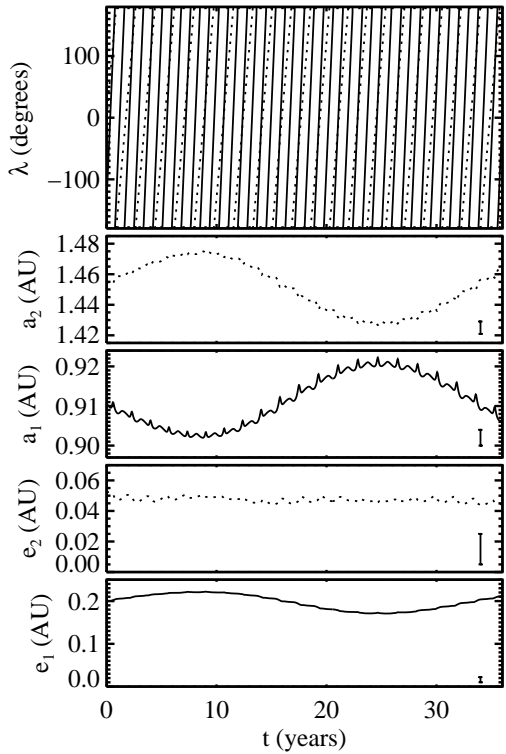}
\caption{  Orbital element changes induced by a mean motion resonance.  The system is the same as in Figure~\ref{fig:sec}, except the outer planet starts at a different period, with a ratio of osculating periods of $2.01:1$.  Note that successive conjunctions (where the lines intersect in the top panel) occur at nearly the same longitude, which causes the period change to grow.  The bar shows the vertical scale of each corresponding panel in Figures~\ref{fig:conjunction} and \ref{fig:sec}, emphasizing that the semi-major axes and eccentricities are experiencing a large oscillation. }
\label{fig:res}
\end{figure}

Next, consider Figure \ref{fig:res}, which is a hypothetical system with periods very close to a ratio of small integers (2.01:1).  In this situation, called a \emph{mean-motion resonance}, conjunctions occur at the same part of the orbit many times in a row, and the change in orbital elements builds.  One may consider a changing period the hallmark of a mean-motion resonance.  Along with period changes come eccentricity oscillations, which can be rather large over only tens of orbits.  For non-coplanar planets, the distance between the location of conjunctions and the intersection of the orbital planes affects their dynamics, so resonances can also involve inclinations and not only eccentricities.  In general, the angles that dictate the behavior of the resonance are called critical angles, and they have the form:
\begin{equation}
\phi = j_1 \lambda_1 + j_2 \lambda_2 + j_3 \varpi_1 + j_4 \varpi_2 + j_5 \Omega_1 + j_6 \Omega_2, \label{eq:dalembert}
\end{equation}
where each planet has a mean motion of $\lambda$, a longitude of ascending node $\Omega$, and a longitude of periastron $\varpi$.  The $j$ values are integers obeying $\sum j_i =0$ (called the \emph{d'Alembert relation}), which is required by the invariance of the system's behavior to the arbitrary reference direction from which angles are measured.  At the very center of each resonance, where the planets come to conjunction at exactly the same point in their orbits, the periods are constant and precession rate is constant.  In this case, there exists a slowly rotating frame in which the motion of each planet is perfectly periodic, yet not perfectly elliptical.

Individual resonances can help keep a system stable.  For instance, when the critical argument for the interior 2:1 resonance has zero libration amplitude ($\theta = 2 \lambda_2 -\lambda_1 - \varpi_1 = 0$, $\dot{\theta}=0$), we may rearrange the equation to read $\lambda_1-\lambda_2 = \lambda_2-\varpi_1$, which shows that when the two planets are at conjunction ($\lambda_1-\lambda_2=0$), then the inner body is also at pericenter ($\lambda_1-\varpi_1=0$), so a close-approach is avoided.  Conversely, whenever the outer body is at the azimuthal location of the inner body's apocenter ($\lambda_2-\varpi_1=\pi$), the two bodies are farthest apart ($\lambda_2-\lambda_1=\pi$).  This argument, applied generally to mean-motions resonances, is called a \emph{resonance protection mechanism} for otherwise unstable systems.

The role and behavior of resonances during planetary migration is beyond the scope of this chapter.  However, the reader is referred to \cite{1976Peale} and \cite{1998Malhotra} for Hamiltonian descriptions of resonances, which can cleanly treat migration.   For a recent applications to exoplanets, see \citep{2002LP} for how two planets can capture into a resonance if their migration converges and \citep{2003C} for how two planets can excite each other's eccentricities as they pass through a resonance while their orbits diverge.

\subsection{ Advanced Interactions }  \label{sec:advanced}

Having surveyed the basic interactions between two planets in \S\ref{sec:types}, we now introduce several more advanced topics.

\begin{figure}
\includegraphics[scale=0.8]{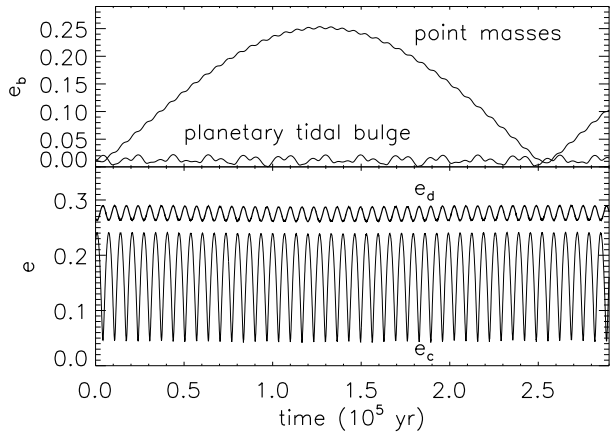}
\caption{  Eccentricities as a function of time for the $\upsilon$~And system.  Orbital elements of \cite{2009Wright}, and edge-on, coplanar orbits were assumed.  We use $m_\star = 1.27 m_\odot$ and ($m_b$, $m_c$, $m_d$)=($0.672$,$1.92$,$4.13$)~$m_{\rm Jup}$.  The outer planets, c ($P_c = 241.33$~d) and d ($P_d = 1278.1$~d), were initialized using Jacobian coordinates, and the plotted eccentricities are also Jacobian.  Two integrations were performed, one in which the planets were point masses and followed the Newtonian equations of motion~(eq.~[\ref{eq:eomnpl}]), and one in which $\mathbf{f}=\mathbf{f}_T$ of equations~(\ref{eq:eom}) and (\ref{eq:planettide}) to model the tidal bulge of planet b ($P_b=4.617136$~d), with $R_p=1.3R_{\rm Jup}$, $k_L=0.34$ assumed.  In the point-mass model, $e_b$ quasi-periodically reaches $\sim 0.25$ (top curve of top panel).  Including the tidal model significantly suppresses the induced $e_b$ (bottom curve of top panel), and including relativistic precession as well (not shown) suppresses it still further.  The evolution of $e_c$ and $e_d$ are almost identical in the two cases, so they are plotted only once.  }
\label{fig:upsand} 
\end{figure}

We previously saw that the eccentricities of planets outside mean-motion resonance can change on a long timescale.  We now extend that concept to systems with three or more planets, systems in which the orbital elements of each planet vary on many different timescales.  Resonances between these timescales can excite eccentricities to very high values (see \citealt{2007Moromartin} for an example in an exoplanetary system).  For the three-planet system Upsilon Andromeda ($\upsilon$~And) in the Newtonian approximation, the innermost planet's eccentricity periodically reaches $\sim 0.25$, compared to $\sim 0.06$ in the absence of the outer planet or $\sim 0.025$ in the absence of the middle planet \citep{2008Barnes}.  Also important in determining the qualitative behavior of the secular dynamics is extra precession, e.g., that supplied by relativistic or tidal effects (\S~\ref{sec:2body}); see \cite{2002WG,2009MGa}.  In Figure~\ref{fig:upsand} we plot the long-term behavior of the eccentricities in the $\upsilon$~And system considering (1) only Newtonian point masses, and (2) an extra force modeling the tidal bulge raised on the inner planet.  The behavior of the outer planets is not much different, but the effect of the extra precession on the inner planet is to detune its pericenter precession rate away from an eccentricity-exciting secular resonance.  Thus, the low current value of $e_b$ argues an additional precession is active \citep{2006AL}; both relativity and tides probably contribute with roughly equal precession rates of $\dot\omega \simeq 10^{-11} s^{-1}$ apiece.  

An extension of secular evolution theory can be made into the regime of high inclination and eccentricity, in which they are strongly coupled \citep{1962K, 1962L}.  In a system with a planet on an initially circular orbit, a third body on a distant exterior orbit will periodically pump the planetary eccentricity to a maximum of
\begin{equation}
e_{\rm max } \approx \sqrt{1 - (5/3) \cos^2 i},
\end{equation}
where $i$ is the mutual inclination which must initially be in the range $39.2^\circ - 140.8^\circ$.  (For a detailed description and a derivation of this behavior, called Kozai oscillations, see \citealt{2007FT}.)  Note that an initially perpendicular orbit ($i = 90^\circ$) leads to an eccentricity of unity.  In some systems, relativistic precession would suppress this behavior, but in others, tidal dissipation would take hold at the eccentricity maximum and circularize the planet at a period of a few days \citep{2007FT, 2007Wu, 2008N}.  An important indicator of whether close-in planets have undergone such an event is their orbital orientation relative to the stellar spin.  With transiting planets, this can be measured via the spectroscopic Rossiter-McLaughlin effect (see \S\ref{sec:kozai} and Chapter 4).

 Another secular effect was found by \cite{2009MGb}, who used a numerical Hamiltonian approach to classify stationary solutions for two-planet, non-coplanar systems.  They found that \emph{interlocking} planetary orbits could stably exist even if no mean-motion resonance protection is active.  Instead, each orbit nodally precesses because of the other orbit, and they do so at the same rate, so the two orbits remain in the same configuration with respect to each other as the whole system precesses.  However, in numerical tests of the full equations of motion, we find that a specific example of this class (fig. 17 of \citealt{2009MGb}) becomes chaotic within several~Myr unless the mass ratios are smaller than $\sim 10^{-4}$.  A system of Neptune-mass planets could thus exist around a solar-like star in this interlocked configuration. 

The high eccentricities of exoplanets call for a mixture of numerical and analytical work to understand the nonlinear dynamics.  The typical method of expanding in small eccentricities and inclinations has been fruitfully supplemented with numerical methods (e.g., \citealt{2004M}) to average over short-timescale effects like those of Figure~\ref{fig:sec}.  Resonant orbits at high eccentricity can have novel properties \citep{2002LP,2003Beauge,2004Lee}.  One way exoplanets can be dynamically simpler than Solar System planets is that they are often hierarchical, i.e., the period ratios can be large.  When this is the case, an expansion in the semi-major axis ratio becomes a useful analytic method (\citealt{2000F} and \S\ref{sec:numtech}).

Mean motion resonances may be shared among more than two planets with critical angles which are extensions of equation~(\ref{eq:dalembert}); these are called three-body mean motion resonances.  They have been shown to destabilize asteroids \citep{1998NM}, the outer planets of the Solar System \citep{1999MH}, and potentially exoplanetary systems \citep{2008Goz3body}.

\subsection{ Chaos } \label{sec:chaos}

	Nothing ensures that planets emerging from a protoplanetary disk---in which collisions were frequent---will be on nearly-Keplerian, stable orbits.  On the contrary, models of planet formation suggest that the orbits of planets should be closely packed together, with the timescale of collisions or ejections being comparable to a system's current age (e.g., \citealt{2000L}).  Indeed, observed multiplanet systems are often close to instability \citep{2008BGR}.  A dynamical system is said to be chaotic if trajectories that are initially separated by an infinitesimal amount diverge exponentially.  This ``microscopic'' definition is often used as a computational tool to detect whether and when ``macroscopic'' events (i.e., ejections and collisions) are likely to occur \citep{1992L,1995MF}, which usually occur on a timescale orders of magnitude longer.  
		
	Resonance overlap is a general condition for strong chaos \citep{1979C}.  Qualitatively, each resonance allows the system to explore a finite zone of semi-major axes, called the width of the resonance.  If two resonances overlap, the system has access to a wider swath of semi-major axes.  This region might even connect to infinity, allowing an ejection.  \cite{1980W} found that resonance overlap could account for the orbits of massless particles becoming chaotic near a planet on a circular orbit.  For a planet with a semi-major axis of $1$ and a mass (relative to the star) of $\mu$, he found that for semi-major axes within $|\Delta a|/a < 1.3$~$\mu^{2/7}$ of the planet, a particle's orbit is chaotic: its fate is either a planetary collision or an ejection.  The mass scaling of this limit is found by comparing the widths of the resonances to their spacing.
     
     	\cite{1989D} presented a complementary understanding of the chaotic zone.  Suppose that two planets follow Keplerian orbits except at their mutual conjunction, where they give each other a kick.  The kick leads to a change in eccentricity which is second-order in the orbital separation, and a change in orbital period which is even higher order.  If the kick in orbital period is strong enough such that the next conjunction occurs more than half an orbital period earlier or later than it would have without the kick, the result of successive kicks is uncorrelated and the system executes a random walk, eventually leading to orbital instability.  This argument leads to nearly the same scaling law of the chaotic region: $|\Delta a| \lesssim 1.24$~$\mu^{2/7}$.  This concept can be readily extended to N-planet systems.  \cite{1996C} applied it to terrestrial planet formation (rather small $\mu$ values), found that there is  no critical separation beyond which stability is assured, and mapped out numerical timescales to instability as a function of orbital separation.  \cite{2007Z} extended these arguments to the mass ratios of most known exoplanets ($\mu \sim 10^{-3}$), finding that kicks at conjunction can explain the instability of planetary systems, but that empirical corrections are needed for quantitative agreement.  
         
         At the relatively larger masses of most known exoplanets ($\mu \sim 10^{-3}$), both these methods start breaking down: (1) their chaotic zones extend to low $p$ among the $(p+1):p$ resonances, so the peculiarities of individual strong resonances need to be taken into account in the resonance overlap picture, and (2) the physics of interactions between planets is not nearly as localized to conjunctions, and this derivation, based on the limit $\mu \rightarrow 0$, breaks down.  Interestingly, the method of resonance overlap no longer breaks down when the concept is applied to even higher mass ratios: \cite{2008M} has shown that for comparable-mass triple stars, $n:1$ resonances are the only ones relevant for stability.  This method has been used to find the instability boundary for planets orbiting one star of a binary star system \citep{2006MW}.
         
         A more general framework for the stability of three-body systems is called Hill stability, in which the orbits of two planets can never cross if a particular inequality is satisfied \citep{1982MB,1993G, 2004VA}.  There is no known sufficient condition for \emph{instability}, unfortunately, but \citealt{2007BGres} have shown that Hill's stability criterion and practical stability are rather close, numerically.
         
         Chaotic trajectories are particularly important during planet formation, in which a given surface density of material is converted into fewer numbers of larger bodies.  Chaotic zones around a given body scale as only a shallow function of its mass, so by putting the same surface density into more widely spaced bodies, the system becomes more stable.  This concept underlies the derivation of the isolation mass --- the mass at which a protoplanet has cleared its feeding zone and no longer accretes --- laid out in \S2.4 of chapter 13 of this volume.
         
      One popular application of chaotic trajectories is the origin of eccentricities.  Exoplanets tend to have larger eccentricities than the Solar System planets.  Perhaps they typically start in systems of several planets of comparable mass, which perturb each other into crossing orbits, resulting in ejection or accretion \citep{1996RF, 2008Chat, 2008JT}.  If indeed this mechanism explains the observed distribution, as discussed in \S3.4 of chapter 9, it implies that most systems of giant planets will self-destruct and many free-floating planets exist.  See \cite{2008FR} for a thorough review of the work on this hypothesis and references to competing hypotheses for the origin of eccentricities.
         
\subsection{ Stable orbits near planets and in habitable zones } \label{sec:otherconfig}
         
     We have seen in \S\ref{sec:chaos} that close to a planet, orbits become unstable.  However, there are particular configurations that allow for long-term stability.  Although none of the following configurations have been observed yet in exoplanetary systems, it is worthwhile to discuss here their existence (their potential detectability is discussed in \S\ref{sec:ttv}).  For each type of configuration, we consider stable orbits of test particles in orbits similar to a planet, then extend the notion to a pair of planets.
     
     	Particles could orbit the planet on the orbit of a satellite, within its Hill sphere.  Consider a particle between the star and a planet on a circular orbit, in a frame that corotates with the planet.  The Hill sphere is the region within which the gravity from the planet ($G m_p r_p^{-2}$, where $r_p$ is the distance to the planet) is comparable to the tidal gravity from the star ($2 G m_\star r_p a^{-3}$, in the radial direction).  Accounting for a differential centrifugal force ($G m_\star r_p a^{-3}$) yields the customary definition for the Hill radius:
\begin{equation}
r_{\rm H} = a \Big(\frac{m_p}{3 m_\star}\Big)^{1/3}. \label{eq:rh}
\end{equation}
Bodies orbiting within some fraction of the Hill radius orbit stably as satellites \citep{2006DWY}.  The same types of orbits exist even if the satellite is massive, even up to the mass equal to that of the planet.  In this case, the planetary mass $m_p$ used to define the Hill radius (eq.~[\ref{eq:rh}]) would be the sum of the two bodies' masses \citep{1986HP}.

     Orbits can also stay near the planet if they are in a 1:1 resonance, of which there are three kinds.  
     
     The first is an extension of satellite orbits to outside the Hill sphere, called quasi-satellite orbits (e.g., \citealt{2008ST}).  From the planet's perspective, the particle would look like a very distant satellite, orbiting the planet in a direction retrograde to their common motion around the star.  From an astrocentric perspective, the particle's orbital phase is similar to the planet's orbit, but its eccentricity is different, which carries it between the planet and the star at periastron and to the far side of the planet from the star at apastron.  If that body is endowed with mass as a second planet, then both orbits will evolve due to their mutual perturbation; the eccentricity can be passed back and forth between the planets \citep{2002LC}.  \cite{2009H} have explored the relationship between this type of 1:1 resonance to satellite-type mutual orbits.
     
     The second type of 1:1 resonance is a Trojan orbit, named after the asteroids that inhabit this resonance with Jupiter.  Their average position is $60^\circ$ ahead of or $60^\circ$ behind the planet in its orbit; these are the stable Lagrange points labeled $L_4$ and $L_5$ \citep{2004D}.  However, the orbits may also wander stably around those points, tracing out a shape of a tadpole, so they are sometimes called tadpole orbits.  Second planets of any mass relative to the more massive planet can exist in these points, but $\mu = (m_1+m_2)/m_\star$ must not exceed $\sim0.038$ \citep{2002LC}, or the system will be unstable.  In this case, both planets will trace out tadpole shapes in the frame rotating with the long-term mean angular velocity, with the size of the tadpole inversely proportional to each planet's mass.
     
     The width of the stable tadpole region scales as $\mu^{1/2}$, which is steeper than the Hill sphere's scaling $\mu^{1/3}$ (eq.~[\ref{eq:rh}]), so for low planetary masses there is a region between them.  In this region lies the third type of 1:1 resonance: the horseshoe orbits.  Such orbits trace out a horseshoe shape in the frame rotating with the orbit of the massive planet, which encompass both $L_4$ and $L_5$.  As with the other resonances, any relative mass of two planets may be in this resonance.  In fact, the Solar System furnishes an example of a $\sim4:1$ mass ratio in this resonance: the Saturnian satellites Janus and Epimetheus \citep{1981DM}.

       In systems with known giant planets, the orbital stability of hypothetical terrestrial planets has been studied extensively, particularly in the region that allows for habitable climates (see chapter 16 regarding habitable zones).  An appropriate approximation is that the terrestrial planet is too small to affect the orbits of the giants: they are treated as test particles (e.g., \citealt{2007Sandor}).  Habitable planets might also reside in the dynamical niches described above: satellites \citep{1997W} or Trojans \citep{2004D}.  In systems already known to have multiple giant planets, the zones of stability can be quite complicated, and numerical integrations are indispensable.
       
       Terrestrial planets that avoid ejection may still be subject to oscillating orbital and spin properties (see \citealt{2003MT} and \citealt{1993LR}, respectively), and these oscillations may cause climate changes (a much-enhanced form of Milankovitch cycles; \citealt{1976HIS}).  The planet may not need to stay in the habitable zone for all of its orbit, depending how well the atmosphere buffers seasonal temperature changes \citep{2002WPproc}, so the upper limit on a habitable eccentricity is a function of planetary properties.  It has been argued that moons of planets with eccentricities as high as $0.69$ (16 Cyg B b) might still be habitable.  Therefore, ejection may be the only dynamical effect that will spoil a habitable world.

\section{ HIGHLIGHTS: DYNAMICS IN NATURE } 
\label{sec:highlights}
     \subsection{GJ 876 and Mean Motion Resonances} \label{sec:gj876}
     	The only exoplanet system hosted by a main sequence star for which non-Keplerian motion has been conclusively detected is GJ 876 \citep{2001Marcy, 2001LC, 2001RL}.  The M-dwarf primary hosts three planets, whose properties are listed in Table~\ref{tab:gj876}.
	
\begin{deluxetable}{lccc}
\tablecaption{The GJ876 System\label{tab:gj876}}
\tabletypesize{\normalsize}
\tablewidth{0pt}
\tablehead{
\colhead{planet} &
\colhead{$m_p$} &
\colhead{ $P [{\rm days}]$ }&
\colhead{ $e$ }
}
\startdata
b&$2.530\pm0.008 m_{\rm Jup}$&$60.83(2)$&$0.0338\pm0.0025$\\
c& $0.790\pm0.006 m_{\rm Jup}$&$ 30.46(2)$&$0.2632\pm0.0013$\\
d&$7.53\pm0.70 m_{\rm Earth}$&$1.93774(6)$&$0.0 ({\rm assumed})$
\enddata
\tablecomments{ Three-planet Newtonian solution from \citep{2005Rivera}, which optimally fits the radial velocities (fig.~\ref{fig:gj876}).  Uses $G m_\star = 0.32 G m_\odot$.  \vspace{0.2 in}
}
\end{deluxetable}

\begin{figure*}
\includegraphics[angle=-90,scale=0.61]{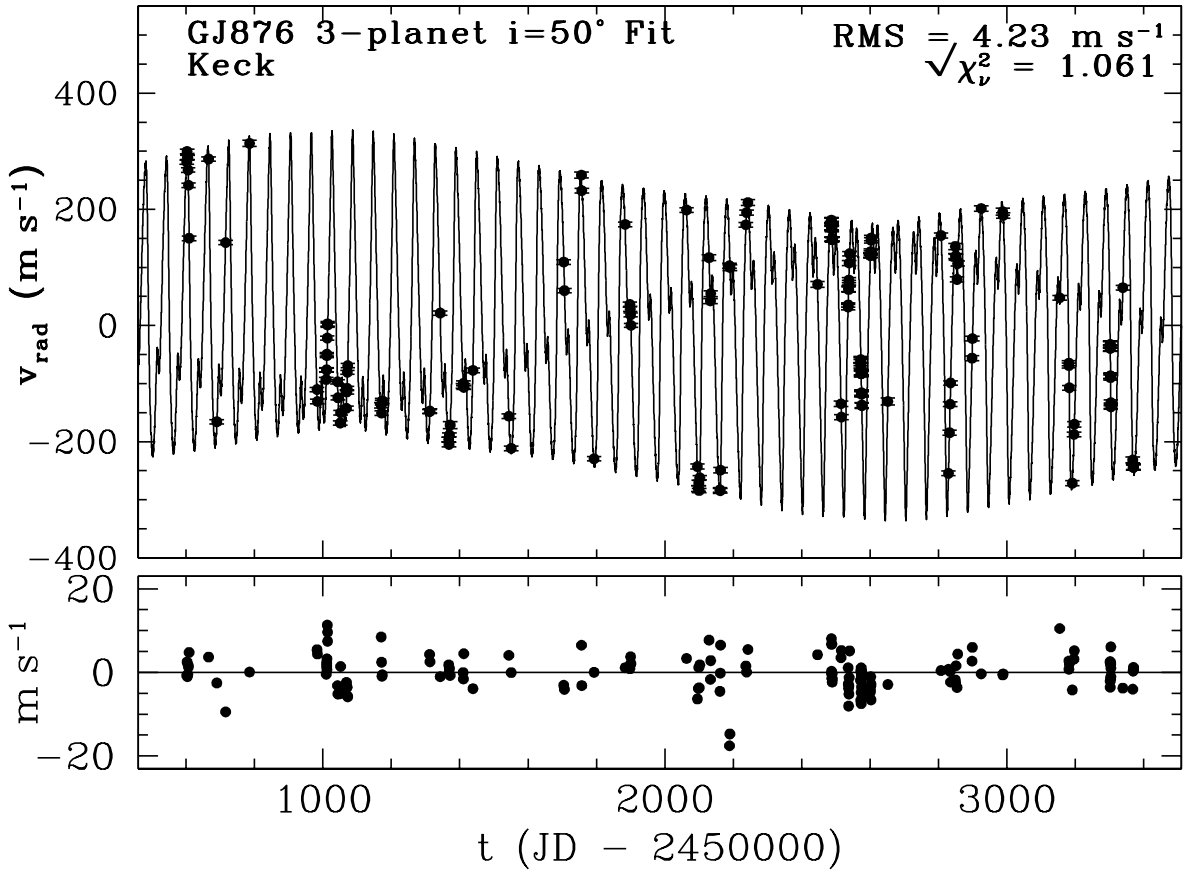}
\caption{  Radial velocity data and best three-planet Newtonian fit to the planetary system GJ876.  From \cite{2005Rivera}.  }
\label{fig:gj876}
\end{figure*}

	These numbers come from a self-consistent Newtonian fit of the radial velocity data (Fig.~\ref{fig:gj876}), in which  a coplanar configuration is assumed and the best-fitting common sky-inclination is found to be $i=50^\circ$ \citep{2005Rivera}.  For typical planets discovered by radial velocity, which are not found to transit, the data are only sensitive to the gravitational influence of the planet in the line-of-sight direction, which yields the quantity $m_p \sin i$, not $m_p$ or $i$ independently.  In this system, the periastron of each planet rotates at a rate proportional to the mass of the other planet, and that rate is well determined by the data.  Therefore $m_p$ is independently measured, breaking the $m_p \sin i$ degeneracy to yield $i$.
	
	However, in this system the dynamically-derived inclination is in tension with the astrometric orbit of planet b \citep{2002Benedict}.  Nevertheless, \cite{2009BS} have fit both datasets simultaneously, and they even determined a mutual inclination of $\Phi_{bc} = 5.0^\circ$~$^{+3.9^\circ}_{-2.3^\circ}$, close but marginally inconsistent with coplanar.  %$\Phi_{bc} = 5.0^\circ$\mbox{$+3.9^\circ$//$-2.3^\circ$}.  
	
	Almost no dynamical constraint can currently be given on the orbital orientation of planet d, from either stability considerations or fits to radial velocity or astrometry data.  Its short period means it is only weakly coupled to the outer planets on the timescale of the data.
	
	The dynamically interesting aspect of the system is that the outer two planets are deeply engaged in a 2:1 resonance, with small-amplitude libration of both critical arguments $\theta_1 \equiv \lambda_c - 2 \lambda_b - \varpi_c$ and $\theta_2 \equiv \lambda_c - 2 \lambda_b - \varpi_b$ about $0^\circ$ (compare eq.~[\ref{eq:dalembert}]).  As a consequence, $\Delta \varpi \equiv \varpi_c-\varpi_b = \theta_2-\theta_1$, also librates around $0^\circ$.  The 3-planet, Newtonian, coplanar model with $i=50^\circ$, for which system parameters were quoted above, yields $|\theta_1|_{\rm max} = 5.4^\circ \pm 0.9^\circ$, $|\theta_2|_{\rm max} = 19.5^\circ \pm 3.8^\circ$, and $|\Delta \varpi|_{\rm max} = 19.4^\circ \pm 4.3^\circ$ \citep{2005Rivera}.
	
	Apart from GJ 876, about 7 other planetary systems have been shown to be in resonance (Table~\ref{tab:res}).  This is usually accomplished, for planets of lower quality data or fewer dynamical times, by noticing the system would be unstable if not for the resonance (\citealt{2005Vogt, 2005Correia, 2006Lee, 2008FM}; for an overview, see \citealt{2007BGres} and Table~\ref{tab:res}).  This logic is needed because (1) not enough orbital periods have been observed for them to be measured to a precision such that non-resonant systems are inconsistent with the radial velocity data, and (2) usually only a small fraction of the resonance libration cycle, or precession cycle, is completed during the observation span, so the orbits appear Keplerian.   

\begin{deluxetable}{lcccccc} \label{tab:res}
\tablecaption{Resonant Systems}
\tabletypesize{\normalsize}
\tablewidth{0pt}
\tablehead{
\colhead{system} &
\colhead{planets} &
\colhead{resonance} &
\colhead{ $P_{in}$ [days] } &
\colhead{ N inner orbits observed } &
\colhead{ $m_{p,\rm{in}}, m_{p,\rm{out}} [m_{\rm Jup} / \sin i]$ }&
\colhead{ Ref. }
}
\startdata
HD 45364 &b-c &3 : 2 & 227&7 & 0.19, 0.66 & A \\
GJ 876 & c-d &2 : 1  & 30& $\sim 80$ & 0.56, 1.94  & B\\
HD 82943 & b-c  &2 : 1  & 441.2 & 8 & 2.01, 1.75  & C \\ 
HD 128311 & b-c  &2 : 1 & 448.6 & 6 & 2.18, 3.21  & D \\
HD 73526 & b-c  &2 : 1 & 188.3 & 13 & 2.9, 2.5 & E \\ 
HD 160691 = $\mu$~Arae & d-b &2 : 1 & 310.5 &9 & 0.52, 1.68  & F \\ 
HD 60532 & b-c &3 : 1 & 201 & 4.5 & 3.15, 7.46 & G\\ 
HD 202206 & b-c  &5 : 1 & 255.87 & 8 & 17.4, 2.4 & H\\ 
\hline
\hline
HD 108874 & b-c &4 : 1  & 395 & 6 & 1.36, 1.02 & I\\  
55 Cnc & b-c &near 3 : 1 & 14.65 & $\sim300$ & 0.82, 0.17 & J
\enddata
\tablecomments{ References: A: Surrounded by chaos \citep{2009C}.  B: King of the resonant planets (\S\ref{sec:gj876}). C: \cite{2004Mayor,2005F,2006Lee,2006GK,2008B},  D: \cite{2005V,2006GK}.  E: \cite{2006T}.  F: also known as HD 160691; \cite{2007P,2008Short}.  G: \cite{2008D, 2009LC}. H: \cite{2005C}. I: not necessarily resonant \citep{2005V}.  J: according to \cite{2008Fischer}, it is actually just outside this resonance. }
\end{deluxetable}

     \subsection{Planets around Pulsar 1257+12}  \label{sec:pulsar}
     The discovery of a planetary system around a pulsar (see chapter 8) preceded the first solid radial velocity detections around main sequence stars \citep{1992WF}.  The discovery technique was to infer the gravitational influence of the planets through tracking the line-of-sight motion of the host star, as in the radial velocity technique.  In this case, however, the light time effect delays pulse profiles.  Nevertheless, with only Keplerian motion detected, the typical $m_p \sin i$ degeneracy held sway. 
      
\begin{figure*}
\includegraphics[scale=1.0]{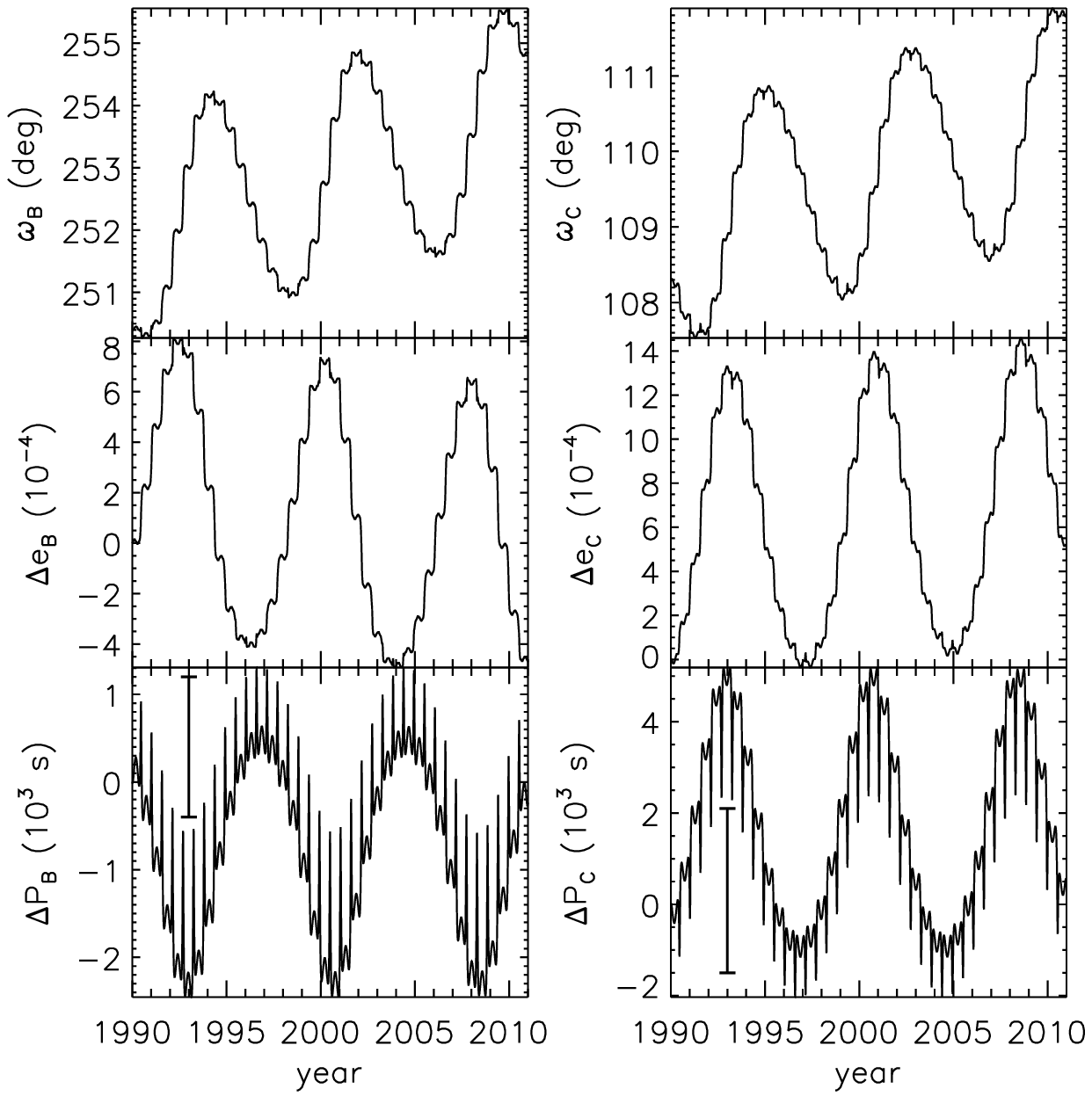}
\caption{  Evolution of the orbital elements of planets B and C around Pulsar 1257+12.  The latest orbital elements \citep{2003Konacki}, taking the planets to be coplanar, are used, and planet A is not included.  Following \cite{1992Rasio}, the elements displayed are argument of pericenter ($\omega$), change in eccentricity ($\Delta e$), and change in period ($\Delta P$) of the Jacobian coordinates of each planet.  Error bars of each planet's $P$ from the discovery paper \citep{1992WF} are indicated in the bottom boxes, showing period changes were nearly detectable (but the changes in $\omega$ and $e$ were far from detectable).  With equally precise data spread over three years, the period variations were detected \citep{1994Wolsz}.  }
\label{fig:pulsarelem}
\end{figure*}

      	The theorists set to work quickly, showing that (a) the perturbations could be detected with more data, (b) they would confirm the planetary nature of the timing residuals, and (c) the amplitude and the character of the detected perturbations would determine the masses of the objects \citep{1992Rasio,1992Malhotra}.  The final point is analogous to how the $m_p \sin i$ degeneracy is broken for the GJ 876 system (\S\ref{sec:gj876}).  However, here the main observable effect was shown to be period variations \citep{1993Malhotra, 1993Peale}, which build up to a sinusoidally varying phase shift with a period of $2\pi/[n_B-(3/2)n_C] \approx 5.5$yr (see Fig.~\ref{fig:pulsarelem}).  The same type of near-resonant behavior is famously active in the Solar System between Jupiter and Saturn, which has a large effect on their orbital phases known classically as the Great Inequality.
     
\begin{figure*}
\includegraphics[angle=-90,scale=0.61]{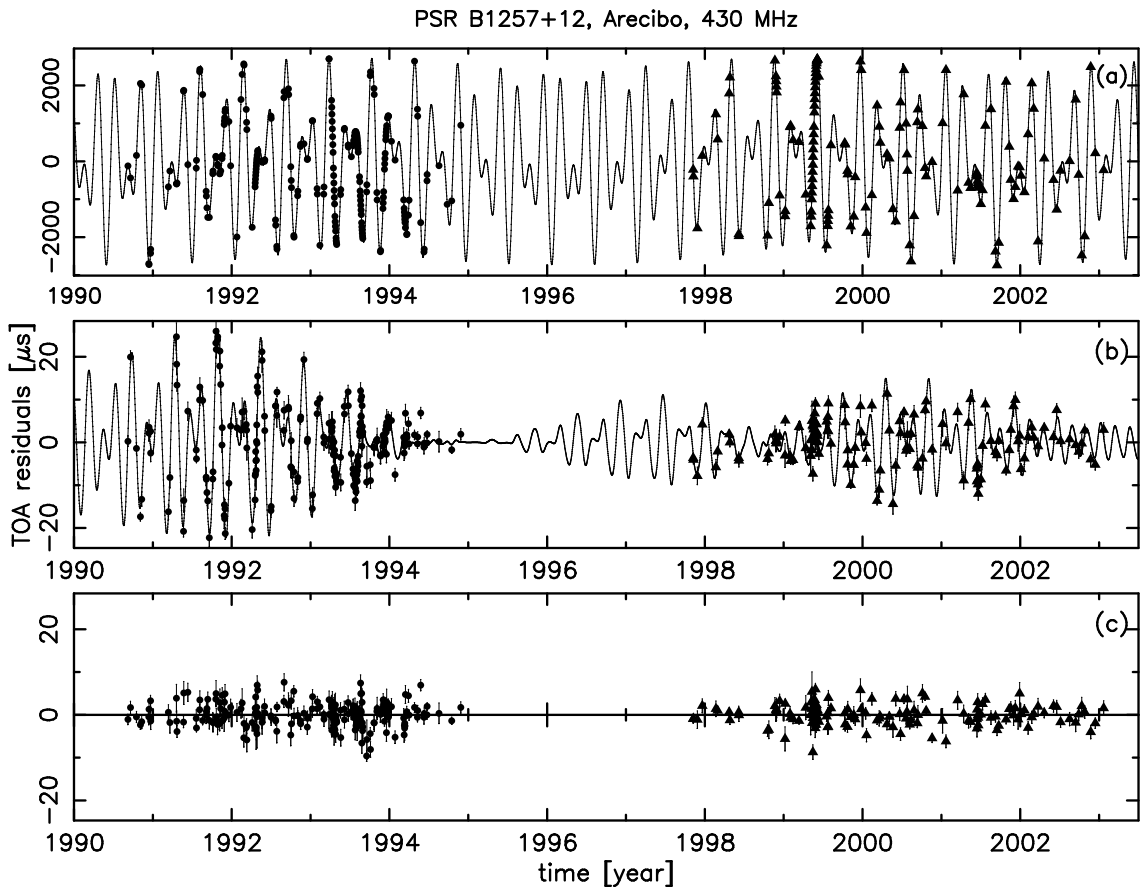}
\caption{  Time-of-arrival (TOA) of pulses for the pulsar that hosts a three planet system.  (a) The daily-averaged data with the best triple-Keplerian model as a solid line; (b) the residuals of the Keplerian fit and the difference between the best Newtonian and Keplerian models as a solid line, (c) the residuals from the Newtonian model.  From \cite{2003Konacki}.  }
\label{fig:psr}
\end{figure*}

	The observers answered the challenge (see chapter 8), not only detecting perturbations to constrain the masses of the planets \citep{1994Wolsz}, but showing their orbits are consistent with coplanar to within $\sim 13^\circ$ \citep{2003Konacki}.  The data and the timing model are shown in Figure~\ref{fig:psr}.

     \subsection{Secular Apsidal Alignment} \label{sec:seclock}
     Eccentricities oscillate and apses precess due to their planets' secular interaction.  If two planets have apses precessing at the same rate, on average, they are said to be in \emph{apsidal lock}, with a critical angle $\Delta \varpi \equiv \varpi_1 - \varpi_2$.  This angle can librate around either $0^\circ$ (apses aligned) or $180^\circ$ (apses anti-aligned), depending on the masses and initial orbital elements, with the restoring torque supplied by the secular terms.  (Mean-motion resonance terms can also result in libration of $\Delta \varpi$, e.g., \citealt{2003Beauge} and \S~\ref{sec:gj876}, but this phenomenon is not our current focus.)  If one planet periodically reaches $e=0$, we follow \cite{2008Ford} in calling such a system \emph{borderline}:  for such systems, $\Delta \varpi$ is on the border between librating and circulating.  We note that there is little dynamical significance to this border, and systems on either side of it remain close to each other in phase space.  In polar coordinates, $(e\cos\Delta\varpi,e\sin\Delta\varpi)$, the difference is just whether the trajectory contains the origin or not. 

   Recently, there have been several attempts to use the libration amplitude, or the proximity to the borderline state, to shed light on earlier epochs in multiplanet systems.  The idea is that if all the planets of the system start out in circular orbits, but one is forced to an eccentric orbit, the system will have a small libration amplitude if this forcing is much slower than the secular timescale \citep{2002CM}, or will be left near the borderline state if this forcing is much faster than the secular timescale \citep{2002Mal}.  The agent imparting the initial eccentricity might be the protoplanetary disk or an additional planet, which is subsequently ejected.  \cite{2005Ford} presented the borderline behavior of the more massive planets in $\upsilon$~And (see Fig.~\ref{fig:upsand}) as evidence of the latter.  Later work showed that, while borderline behavior is surprisingly common among multiplanet systems \citep{2006BGb}, it is perhaps even \emph{too} common for simple models of scattering among planets to be the explanation \citep{2007BGaps}.  The final answer awaits a rigorous statistical comparison between the secular behavior resulting from scattering simulations (e.g., \citealt{2008Chat}) and the secular behavior inferred from data (e.g., \citealt{2009V}).
    
     \subsection{Kozai oscillations} \label{sec:kozai}
     One of the first very eccentric exoplanets, 16 Cyg Bb, caused excitement because its orbit is quite unlike the nearly circular orbits of the giant planets of the Solar System, and it was unclear how a giant planet could form on an eccentric orbit.  It also suggested that the low eccentricities of the previous discoveries were perhaps not primordial, but tidally damped.  This particular system has a distant companion star, which could cause Kozai eccentricity oscillations (\S2.3, \citealt{1997Holman,1997Mazeh}), resulting in the high eccentricity.  
     
     As pointed out by \cite{2008Takeda}, the four planets with the highest eccentricities ($e>0.8$), and $\sim 50\%$ of the $18$ planets with $e>0.6$, have confirmed stellar companions; this correlation is statistically significant, considering that only $\sim 16\%$ of exoplanet hosts have stellar companions \citep{2008MN} due to survey biases.  These numbers may be interpreted as a statistical detection of Kozai oscillations, which have much too long a timescale to be directly observed in the current data of any particular system.  However, there is additional evidence that one of these four systems, HD 80606b with e=0.93, has indeed experienced Kozai cycles.  It was shown that a natural prediction of the Kozai scenario (as described by \citealt{2003WM}) is that the stellar spin is currently misaligned from the planetary orbit \citep{2007FT}.  This prediction of misalignment was recently verified \citep{2009M,2009W}.

\section{FUTURE PROSPECTS} \label{sec:future}
     \subsection{ Searching for small planets } \label{sec:ttv}
     
     One of the applications of dynamical calculations, which has yet to be realized, is the detection of previously unknown small planets via their dynamical effect on known planets.  This potential is particularly ripe for the transit timing method.
     
     Transiting planets offer a wealth of information that is inaccessible for a usual radial-velocity-detected planet (see chapter 4).  An exciting opportunity is afforded by the extreme phase sensitivity of a transit light curve.  With high-quality data, a timing precision of $\sim10 s$ is achievable, which translates to a phase measurement of $\sim 0.01^\circ$ along a 4-day orbit.  Only very slight perturbations lead to along-track variations of that order.  In comparison, for a Jupiter-mass planet on a 4-day orbit around a solar-mass star, a radial velocity datum with precision of 3~m~s$^{-1}$ translates to a phase measurement of $\sim 1^\circ$.
     
\begin{figure}[t]
\includegraphics[scale=0.8]{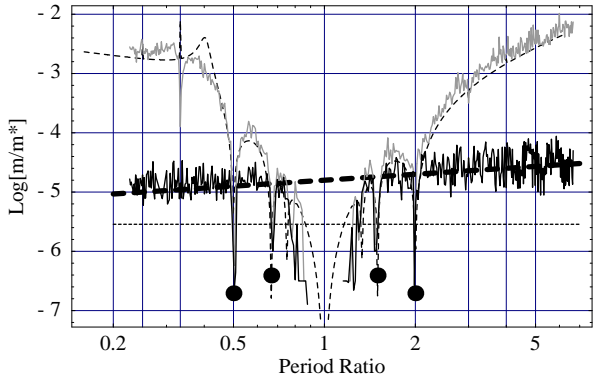}
\caption{  The 3-$\sigma$ upper limits on the mass of a hypothetical second planet (with assumed eccentricity $0.02$) relative to its host star in the HD 209458 system, based on the precise and non-variable transit times of planet b (solid, gray curve) and that data combined with the radial velocity time-series (solid, black curve).  A perturbation theory calculation (thin, dashed curve) matches the transit time constraints between resonances, and an analytic expression (large dots) matches within resonances.  The most sensitive constraints are within resonances, and extend an order of magnitude lower than both upper limits from radial velocity (thick dashed line) and the mass of the Earth (thin dashed horizontal line).  Near the period ratio of 1, orbits of small bodies are within the chaotic zone (\S\ref{sec:chaos}) and are thus unstable.  From \cite{2007AS}.  }
\label{fig:as07}
\end{figure}

     Though it has potential, no compelling examples of transit time variations have yet been published.  However, \cite{2005SA} --- and many authors following their example --- have published stringent upper limits on hypothetical second planets, which are required by constant transit times.  \cite{2007AS} have analyzed transits of the first-discovered transiting planet HD 209458b using the {\slshape Hubble Space Telescope}, searching for other companions.  They were able to put stringent upper limits on the mass of a hypothetical second planet, as a function of its period: see Figure~\ref{fig:as07}.  The sensitivity of transit time measurements is orders of magnitude better within resonances than midway between them \citep{2005A,2005HM}.  This is simply a restatement of our earlier identification of resonances as effective locations for $a$ and $P$ variations.  Thus they provide complementary information to radial velocity observations, which are not particularly good at discovering planets in resonance (especially the 2:1 resonance; \citealt{2010ALC}), but they have sensitivity over a wide range of periods.
     
   Although mid-transit times are the most sensitive characteristic to period changes over timescales short compared to the observations, transit \emph{duration} variations (TDV) have been recognized as more sensitive to variations of much longer timescale \citep{2002M, 2007HG, 2008PK}, and they may eliminate degeneracies inherent to TTV measurements \citep{2009K}.
    
    Transit timing measurements are a possible way to find objects in qualitatively different orbits than those available to the radial velocity method.  Trojans are hard to pull out of the radial velocity data because their orbits have the same harmonics as the main planet \citep{2006GK}.  However, if the two planets have low eccentricity and a large libration amplitude, the signal would be a single sinusoid with an amplitude that slowly oscillates (every $\sim 10$ orbital periods), markedly different from the signal of a single planet \citep{2002LC}.  Such a libration could be easily seen in transit data \citep{2007FH}.  Trojan planets that make a perfect equilateral triangle with the star would not librate at all.  However, a combination of transit and radial velocity data could still detect it \citep{2006FG}, which has been used to place upper limits of varying sensitivity in 25 systems \citep{2009MW}.  Moons of giant planets could also be searched for \citep{2007SSS, 2009K} by transit timing, but the conceivable limits are well above the mass of Earth's moon or of the moons of giant planets in the solar system.  The masses of moons may be limited by their formation mechanism \citep{2006CW} or by the requirement that tidal evolution as not destroyed them \citep{2002BO,2009Cass}.  Nevertheless, many of the giant exoplanets are in the habitable zone of their stars, so the first such planet to transit will be carefully scrutinized for both habitable moons \citep{1997W} and habitable trojan companions \citep{2004D}.

     \subsection{ System architectures through detecting perturbations } \label{sec:arch}
     
     In this section we shall examine how more observations of non-Keperian motion will contribute to our understanding of the architectures of planetary systems, some aspects of which we currently know very little.  We shall see that (a) resonant orbits and (b) mutual inclinations of planetary systems will likely enjoy fundamental advances in the coming years via detected perturbations.
     
     First, a historical look at the two well-observed non-Keplerian systems shows their enormous contribution to our understanding of planets.  In 1992, detecting perturbations for the planets of PSR 1257+12 was decisive in demonstrating the orbital and planetary nature of the signal, rather than an unforeseen pulsar oscillation \citep{1992Rasio, 1994Wolsz}.  Thus the confirmation of the first exoplanets, and the detection of the only sub-Earth-mass exoplanet known, required the modeling of non-Keplerian motion.  In 2001, detecting perturbations for the planetary system GJ876 was important for demonstrating the true masses of planets orbiting main-sequence stars; they are smaller than masses of brown-dwarfs \citep{2001Marcy}.  Only later did true masses become known for many planets thanks to transit measurements.  This history suggests we can expect perturbations to reveal new aspects of planetary systems.
     
     	One aspect of system architectures that is ripe for observational input is resonant orbits.  The resonance of GJ876bc has already been mapped out in detail, but orbital changes due to resonance could be detectable on decade timescales in other systems as well \citep{2005F,2005Correia,2009C}.  It might seem that 55 Cnc, with 5 planets, two of which are close to resonance, would be a good candidate.  But neither \cite{2002Marcy} nor \cite{2008Fischer} found improved fits when the system was modeled with Newtonian equations (\ref{eq:eomnpl}) rather than independent Keplerians.  Though perturbations have not been detected in most systems, see Table~\ref{tab:res} for a summary of the systems that have been plausibly claimed to be resonant.  Determining whether this conclusion is robust, on a case-by-case basis, and finding the libration amplitude, has not received much observational attention.  However, the frequency of resonances and the expected libration amplitude of the critical angles, or their circulation, has been a frequent topic of theoretical research.  Currently observed resonances have the potential to constrain nebular conditions that lead to their formation and/or destruction \citep{2002LP, 2003Beauge,2008Adams, 2009L}, perturbative events in the system's history \citep{2006SK, 2008RP, 2009Lee}, and tidal dissipation for close-in inner planets \citep{2003Novak,2007TP}.  Therefore, with some additional high-quality observational input, we stand to learn fundamental information about the architecture of resonances, which in turn informs several classes of theories.

Just outside of resonance, an oscillation in the orbital periods and eccentricities is expected, as shown in \S\ref{sec:pulsar}.  Depending on the proximity from resonance, it may be small enough to have eluded detection even in well-studied systems, e.g., in 55 Cnc.  Conversely, since no perturbations are detected there, a constraint may be placed on the true masses, as resonance widths grow with mass \citep{1992Malhotra}.  Thus the \emph{absence} of detected non-Keplerian motion can yield important constraints.

	A second aspect of system architectures that has received rather little attention is mutual inclination.  Although mutual inclination is recognized as a fundamental quantity for planet formation theory (e.g., it was a key inspiration for Laplace's nebular hypothesis), it cannot be measured by radial velocity if planets stay on Keplerian orbits.

Secular precession due to non-coplanarity has been detected in triple stars by radial velocity (e.g., \citealt{2000J}) and by eclipses (e.g., \citealt{2000TS}), and it has led to constraints on their mutual inclination.  In a non-coplanar two-planet system, each orbit exerts secular torques on the other, and they each precess like a top.  The precession period in the low-eccentricity, nearly coplanar case, with $P_{\rm out} \gg P_{\rm in}$ is:
\begin{equation}
P_{\rm sec} = \frac{8 \pi}{3} \frac{P_{\rm out}^2}{P_{\rm in}} \frac{m_\star}{m_{p,{\rm out}} + m_{p,{\rm in}} (P_{\rm in}/P_{\rm out})^{1/3}}, \label{eq:psec}
\end{equation}
where ``in'' and ``out'' refer to the inner and outer planets, respectively \citep[\S7.2]{1999MD}.  This period is also the right order-of-magnitude for moderate mutual inclinations and eccentricities, and it also roughly describes the periapse precession of such a system \citep{2003LP}.  In systems for which this period is small, one might hope to observe precession directly and probe mutual inclination.  The shortest precession period due to secular terms among known systems is that of GJ876bc, with $P_{\rm sec} \gtrsim 100$~yr.  Although this timescale is much longer than a manageable observing program, even a small part of this cycle could produce observable effects, because over the complete cycle the orbit can entirely reorient.  So far mutual inclination in planetary systems has only been measured in two systems, based on resonant perturbations \citep{2003Konacki,2009BS}.

Transit-timing variations on orbital timescales may be a much faster route to mutual inclinations.  Theoretical evidence has been building that the pattern made by transit times can reveal mutual inclination in a single passage of an external, eccentric planet \citep{2003Bork,2009Bakos} or the detailed signal of short-term interactions \citep{2005A,2009NB}.  Therefore, as soon as transit time variations are detected, a constraint can be put on mutual inclination.

The transit-timing method may prove to be crucial in interpreting aspects of the HAT-P-13 planetary system, for which planet b, the inner planet, transits.  According to \cite{2009Batygin}, following \cite{2007Mard}, planet b should have damped to a calculable, non-zero eccentricity due to forcing by the outer planet, which is massive and eccentric.  The value of this forced eccentricity depends on the planet's tidal deformability (the Love number $k_L$) and thus its interior structure.  Thus for the first time, we could have a constraint on the mass distribution interior to an exoplanet.  However, this chain of logic assumes that the two planets are coplanar, so it is crucial to establish that fact.  \cite{2009Bakos} showed in the discovery paper that $\sim 5$~s transit time variations are expected if the system is coplanar, and even larger variations (of a different shape) are expected if the system is non-coplanar.  Therefore, a tidal bulge may be soon inferred, resting on two aspects of non-Keplerian dynamics: (1) orbital timescale perturbations for observers to determine the mutual inclination, and (2) secular timescale and tidal evolution, which causes the interior structure of the planet to feed back on its observable eccentricity.

For further reading, the standard reference textbook for planetary dynamics is {\em Solar System Dynamics} by \cite{1999MD}.  It contains much of the material here, with all the required mathematical detail, but it was written too early to include many results on exoplanets.  Two recent reviews on the dynamics of exoplanet systems fill in many of the details of this chapter: \cite{2007Mich} treat secular and resonant effects in two-planet systems, and \cite{2008Breview} discuss resonant dynamics (particularly of the 2:1 resonance) and orbital fits to such systems.

The study of non-Keplerian motion in exoplanets has played a critical role in understanding their nature and histories, from the first known exoplanets (around a pulsar, of all places) to systems of planets in dynamical configurations which shed light on their early history.  The relatively simple equations hold a wealth of complexity, the extent of which is still being mapped in parallel with the continued discovery of planets unlike those of our Solar System.

\vspace{0.4 in}
\emph{Note added in proof:}
The first haul of candidate multiple-planet systems discovered by the transit method has now been brought in by the \emph{Kepler} mission \citep{2010S}, which suggests timing measurements (\S\S 2.5, 5) will be revolutionized, for several reasons.  First, Kepler is efficient at detecting much longer-period transiting planets than has been possible by ground-based surveys, and the timing signals scale with the orbital periods of the planets (e.g., \citealt{2005A}).  Besides that simple scaling, systems with periods $\gtrsim 10$~days are little affected by tidal dissipation, so pairs of planets can stay in resonance, resulting in very large timing signals their whole lifetime.  Second, in the single-transiting case, various system architectures can yield nearly indistinguishable timing signals \citep{2007FH,2008NM}; but in systems of two planets that both transit and show timing variations, the system parameters (including masses; \citealt{2005HM}) can be much more easily inferred from the data.  Finally, \cite{2010S} showed that their 5 systems with multiple transits will have timing variations readily apparent in the full \emph{Kepler} dataset.  They also showed that for every system with multiple planets that transit, due to viewing geometry there must be 2-20 or more systems like it, each with only one planet displaying transits.  These two facts together imply that \emph{Kepler} will discover tens if not hundreds of planets on detectably non-Keplerian orbits!

%\section{Acknowledgements} 
\acknowledgements
I acknowledge support from the Michelson Fellowship, supported by the National Aeronautics and Space Administration and administered by the Michelson Science Center, and I thank an anonymous referee, E. Agol, J. Bean, R. Dawson, E. Ford, K. Gozdziewski, M. Holman, R. Malhotra, R. Murray-Clay, D. Ragozzine, J. Steffen, S. Tremaine, D. Veras, J. Winn, and J. Wisdom for discussions and comments on a draft of this chapter which considerably improved it.

\bibliography{F09bib} \bibliographystyle{apj.bst}

\end{document}